\documentclass[aps,prl,reprint,showpacs,showkeys,amsmath,amssymb]{revtex4-1}
\usepackage{graphicx}
\newcommand{\fe}{\mathfrak{e}}

\newcommand{\SO}[1]{\text{SO}(#1)}

\begin{document}

\title{A Short Proof of the Reducibility of Hard-Particle Cluster Integrals}
\author{Stephan Korden}
\affiliation{Institute of Technical Thermodynamics, RWTH Aachen
University, Schinkelstra\ss e 8, 52062 Aachen, Germany}
\date{\today}

\begin{abstract}
The current article considers Mayer cluster integrals of $n$-dimensional hard
particles in the $n>1$ dimensional flat Euclidean space. Extending results from
Wertheim and Rosenfeld, we proof that the graphs are completely reducible into
1- and 2-point measures, with algebraic rules similar to Feynman diagrams in
quantum field theory. The hard-particle partition function reduces then to a
perturbatively solvable problem.
\end{abstract}
\maketitle
{\em Introduction ---}
Hard-particle systems are the simplest example of classical fluids developing
a gaseous and solid phase. They are therefore a suitable starting point for
understanding realistic fluids, as the asymptotic limit to high packing
densities is dominated by their geometry and volume dependence. Understanding
the hard-particle free-energy structure is thus an important open problem.
Comparing this to the better investigated statistical and quantum field
theories, one can identify two differences that are unique to classical
particles: their interaction potentials $V(|\vec{r}\,|)$ are strongly singular,
i.e. more divergent than $V(r)\sim r^{n-1}$ in $n$ dimensions, causing the
integral $\int V(r)r^{n-1}dr$ to be infinite over the particle's finite domain.
Furthermore, the interaction is not local, resulting e.g. in the blocking of
particles at high densities. The first problem has been solved by Mayer
\cite{mcdonald}, by expanding the partition function in $f(r_{ij})= \exp{(-\beta
V(r_{ij}))}-1$ and representing it in cluster integrals. Whereas an approach to
the second problem has been found by Rosenfeld \cite{rosenfeld-structure,
rosenfeld-freezing, rosenfeld1, rosenfeld-mixture, rosenfeld-closure,
rosenfeld2} in splitting the f-function into weight functions or 1-point
measures $f(r_{ij}) \sim \mu(\vec{r}_i) \cdot \mu(\vec{r}_j)$. His fundamental
measure theory (FMT) of hard spheres has been extended to fluids of convex
particles \cite{goos-mecke, mecke-fmt} and to the crystalline phase
\cite{tarazona-rosenfeld, tarazona}, see \cite{fmt-review} for a review. The
weight functions entering the FMT, were then further investigated by Wertheim
\cite{wertheim-1, wertheim-2, wertheim-3, wertheim-4}, expanding ring graphs
into 2-point measures. Both approaches make use of the observation of Kihara and
Isihara \cite{isihara-orig, kihara-1, kihara-2} that the second virial
coefficient can be understood as the kinematic fundamental formula of integral
geometry developed by Blaschke, Santalo and Chern \cite{blaschke, santalo-book,
chern-1, chern-2, chern-3}. For further applications and an overview of integral
geometry see also \cite{mecke-phd, bernig-rev} and \cite{mecke-fmt, goos-mecke}
for the discussion of its relation to the 1-point measures in FMT.

In this article we will explain how the intimate relationship between
geometry and symmetry leads to the decoupling of arbitrary Mayer clusters
into 1- and 2-point measures, extending Wertheim's result for ring graphs and
justifying Rosenfeld's ansatz for the free-energy. In the first part, we will
consider the splitting of the clusters into vertices, using their relation to
integral geometry. Whereas the second part focuses onto the process of
rejoining the vertices into 2-point measures. As the arguments are not
restricted to $3$ dimensions, we consider $n>1$ dimensional particles, embedded
into the flat Euclidean space $\mathbb{R}^n$. It will be shown that the 1-point
measures have some similarity to the wave functions of bosonic field theory,
where the 2-point measure can be interpreted as a propagator and the Mayer
cluster as a Feynman diagram, constructed from a finite set of irreducible
vertices. Each loop in the diagram introduces constrains that parallels charge
and momentum conservation. Our discussion will focus on the mathematical
aspects leading to the decoupling. For an extended exposition with applications
to $3$ dimensions we refer to \cite{korden}. 

\begin{figure}
\centering
\includegraphics[width=2.5cm,angle=-90]{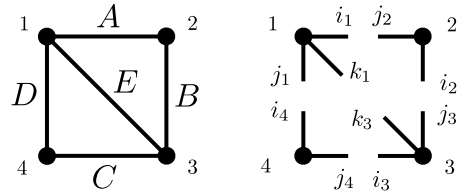}
\caption{The 4-particle cluster integral splits into two 2- and 3-vertices along
the lines $A,\ldots,E$.}
\label{fig:vertex}
\end{figure}

{\em Splitting the Cluster ---}
Mayer clusters \cite{mcdonald} are graphical representations of cluster
integrals over the flat Euclidean space $\mathbb{R}^n$ with bonds as f-functions
and black circles representing space points of overlapping particles. In the
following we will call a black circle with $k$ outgoing lines a $k$-vertex,
corresponding to $k$ intersecting particles $D_1\cap \ldots \cap D_k\neq 0$.
FIG.~\ref{fig:vertex} gives an example for a star graph with 2- and 3-vertices. 
Each diagram is then integrated over all positions and orientations of its
particles in $\mathbb{R}^n$. From an active point of view, this is identical to
an integration over all translations and rotations in space, generated by the
$k$-fold tensor product of the $n(n+1)/2$ dimensional Lie group $\text{ISO}(n)
\simeq \mathbb{R}^n \times \SO{n}$. Now, let $\Sigma$ be a smooth, orientable
Riemannian manifold of $n-1$ dimensions, embedded into the flat Euclidean space
$D:\Sigma\hookrightarrow \mathbb{R}^n$. Each point in the domain $p\in D$
belongs to a local, orthonormal coordinate frame $(p,e_n,\ldots, e_1)$ of the
tangential space. The elements of the dual space then derive from Cartan's
defining equations $dp=e_i\theta^i$ and $de^i = \omega^{i}_{\;j} e^j$
for $i,j=1,\ldots,n$. The vielbeins $\theta^i$ and spin connections
$\omega^i_{\;j}$ are the elements of the Lie algebra $\text{iso}(n)$ with
the Maurer-Cartan equations \cite{helgason} following from the vanishing torsion
$T^i= d\theta^i+\omega^i_{\;j}\wedge\theta_j= D^{(n)}\theta^i = 0$ and the
curvature of the flat Euclidean space $\Omega^{(n)}_{ij} =d\omega_{ij}
+\omega_{i}^{\;k} \wedge\omega_{kj} =0$. Here, $D^{(n)}$ is the covariant
derivative in $n$ dimensions. Let $e_n$ be the outward normal vector of the
surface $\Sigma=\partial D$. The curvature form $\Omega^{(n)}_{ij}$ splits
then into the two contributions
\begin{eqnarray}
\Omega^{(n-1)}_{\alpha\beta} = \omega_{\alpha n}\wedge &&\omega^{n}_{\;\beta} 
\label{gauss}\\
D^{(n-1)}\omega_{n\alpha} = 0 && \hspace{3em}\text{for}\quad
\alpha,\beta=1,\ldots, n-1\;.
\label{gauss-codazzi}
\end{eqnarray}
Now, the first part of the proof will show, by explicit calculation, that the
integral density of the kinematic fundamental formula factorizes into an
infinite set of 1-point measures, analogous to the known integrated form
\begin{equation}\label{split}
\mu^k(D_1 \cap D_2 \cap\ldots) = C^k_{ij} \mu^i(D_1) \mu^j(D_2\cap\ldots)\;,
\end{equation}
where the 1-point measures reduce to the finite basis $\mu_i$ of Minkowski
functionals \cite{santalo-book}.

As a first and illustrative example of integral geometry, consider one particle
moving in $\mathbb{R}^n$. The integration over all rotations and translations
introduces the Haar measure of $\text{ISO}(n)$ as the $n(n+1)/2$ form
$\wedge_{1\leq i<j}^n \omega_{ij}\wedge_{i=1}^n \theta_i$. For Riemannian
manifolds this differential form is identically zero, as the vielbeins and spin
connections are related by $T^i=0$. To get a non-trivial result, observe that
rotations and translations only in the normal direction change the particle's
position in space, reducing the integration to the $n$ dimensional coset space
$\SO{n}/ \SO{n-1}$. With equation (\ref{gauss}), the reduced kinematic measure
can be written as:
\begin{equation}\label{measure}
dD=\wedge_{1\leq \alpha}^{n-1} \omega_{\alpha, n}\wedge\theta_n = \gamma_{n,n-1}
K(\partial D)\wedge \theta_n
\end{equation}
with the Euler class $K$ and a constant $\gamma_{n,n-1}$, determined in
\cite{korden}, but irrelevant for the current discussion. The Euler class is a
topological invariant \cite{bott-tu}, depending only on the surface $\Sigma$ of
the particle and not its domain. Furthermore, it is a local function, invariant
under coordinate transformations of $\SO{n-1}$. 

As already noted before, the vertex of order $k$ corresponds to $k$ intersecting
domains $D_1\cap \ldots \cap D_k\neq 0$ integrated over all translations and
rotations. With $dD_m$ as the reduced kinematic measure of particle $m$, the
vertex contribution to the cluster can be rewritten
\begin{equation}\label{vertex}
\begin{aligned}
& dD_1\wedge \ldots \wedge dD_k 
\left.\right|_{D_1\cap \ldots \cap D_k\neq 0}\\
& = \gamma_{n,k}\,K(\partial (D_1\cap \ldots \cap D_k))\wedge dD_2\wedge \ldots
\wedge dD_k\;
\end{aligned}
\end{equation}
generalizing (\ref{measure}). From this we will now derive two characteristic
properties of the vertices, leading to the splitting of diagrams into 1-point
measures. 

Observe, that the Euler form of (\ref{vertex}) depends only on the boundary of
the intersecting particles. As follows from homology theory \cite{bott-tu},
the boundary operator is a derivation $\partial\{\text{point}\}=0$,
$\partial^2=0$ acting on homology elements. Its operation on two domains
\cite{chern-1}
\begin{equation}\label{boundary}
\partial (D_1\cap D_2)=\partial D_1\cap D_2 + D_1\cap\partial D_2 + \partial
D_1\cap \partial D_2
\end{equation}
can then be extended in an obvious way to an arbitrary set of intersecting
particles. But as each operation of $\partial$ decreases the space dimension by
one, the intersection of $n+1$ surfaces is identically zero. This proofs the
first result that {\em the number of irreducible terms of a vertex is at most
$n$ }, reducing the problem to intersections of the form $\Sigma_1 \cap \ldots
\cap \Sigma_k \cap D_{k+1} \cap \ldots \cap D_m$ for $k\leq n$. This can be
further simplified, as the coordinate frames of intersecting domains $D_{k+2}
\cap \ldots \cap D_m$ are unrelated and decouple as volume forms from
(\ref{vertex}). The case of $\Sigma_k\cap D_{k+1}$ is similar but introduces the
constrain $e^{(k)}_n=\pm e^{(k+1)}_n$ among the normal vectors. Again, the
integral measure of the domain factorizes from (\ref{vertex}) as a volume form,
with a sign absorbed in $\gamma_{n,k+1}$. It remains to determine the kinematic
measure for $\Sigma_1 \cap \ldots \cap \Sigma_k$, generalizing the explicit
calculation of Chern \cite{chern-1} for $k=2$ to proof that (\ref{vertex})
decouples into 1-point measures. First, introduce the orthonormal frames
$(e_1^{(a)}, \ldots, e_n^{(a)})$ for the $a=1,\ldots,k$ particles, related by
the $k(k-1)/2$ intersection angles 
\begin{equation}\label{intersection-angles}
<e_n^{(a)}|e_n^{(b)}> = \cos{(\phi_{ab})}
\end{equation}
These additional constrains generalize the coset space of one particle to the
$kn$ dimensional, reduced kinematic measure of $k$ intersecting surfaces
$\text{ISO}(n) \times \SO{k}/ \text{ISO}(n-k)$. In order to calculate the new
$n-k$ form $K(\Sigma_1\cap\ldots\cap\Sigma_k)$, define the orthonormal frame 
\begin{equation}
\begin{aligned}
(e_1, \ldots, e_{n-k}, & v_{n-k+1}, \ldots, v_n)\\
& =  S (e_1, \ldots, e_{n-k}, e_n^{(k)}, \ldots, e_n^{(1)})\;,
\end{aligned}
\end{equation}
where $S\in \text{GL}(k)$ is a linear, invertible transformation, most easily
constructed by the Gram-Schmidt process. To parameterize the change of angular
phase at the intersections, introduce the additional $k(k-1)/2$ angular
coordinates $0\leq \sigma_{ab}\leq \phi_{ab}$, interchanging the $n-k$ normal
directions $R(v_n, \ldots, v_{n-k+1}) = (\eta_n, \ldots, \eta_{n-k+1})$ for
$R(\{\sigma_{ab}\}) \in \SO{k}$. The Euler class of $k$ intersecting surfaces is
then determined by 
\begin{equation}\label{euler-n-k}
K(\Sigma_1\cap\ldots\cap\Sigma_k) = \wedge_{\alpha
=1}^{n-k} e_\alpha d\eta_n \wedge_{n-k+1=i<j}^n \eta_i d\eta_j\;,
\end{equation}
a $n-k$ form in the common particle coordinates and the $\SO{k}$ measure
of $R(\{\sigma_{ij}\})$. But one unattractive feature of (\ref{vertex})
still remains, as the result is not obviously invariant under particle
permutations. This can be solved by rewriting the connections in the principal
representation $\omega^{(a)}_{\alpha, n} = \kappa_\alpha^{(a)}
\theta_\alpha^{(a)}$ and transforming to the coordinate system of particle
$\Sigma_1$. After integrating out $\sigma_{ab}$, one obtains a polynomial in the
principal curvatures $|K(\{\kappa_\alpha^{(a)}\},\{\phi_{ab}\})|$, depending
only on the intersection angles and the form $\wedge_{\alpha=1}^{n-k}
\theta^{(1)}_\alpha$. To get a symmetric result, we still have to incorporate
the intersection constrains into the remaining integral measure $dD_2 \wedge
\ldots \wedge dD_k$. As explained in more detail in \cite{korden}, a suitable
coordinate system for the domain $D_a$ follows from the above transformation
$(e_1^{(a)}, \ldots, e_{n-a}^{(a)}, S e_{n}^{(a)}, \ldots, S e_n^{(1)})$,
corresponding to the decomposition $\SO{k} = \SO{k}/ \SO{k-1} \times \ldots
\left. \times \SO{3}/\SO{2}\right. \times \SO{2}$. Taking into account the
additional constrains $\theta_n^{(a)}=0$ at $\Sigma_a$, the integral measure for
the $k$-vertex obtains the final form
\begin{equation}\label{vertex-reduced}
\begin{aligned}
& dD_1 \wedge \ldots \wedge dD_k 
\left.\right|_{\Sigma_1\cap \ldots \cap \Sigma_k\neq 0}\\
& = \tilde\gamma_{n,k} |K(\{\kappa_\alpha^{(a)}\},\{\phi_{ab}\})|\, dG\wedge
d\Sigma_1\wedge \ldots
\wedge d\Sigma_k\;,
\end{aligned}
\end{equation}
factorizing into the reduced kinematic measures of individual particles
$d\Sigma_m$, the group measure $dG$ and the determinant $|K|$.

Equation (\ref{vertex-reduced}) generalizes Wertheim's decoupling of the second
virial diagram \cite{wertheim-1} into 1-point measures $\mu^i(D_a)$ by
expanding the $\phi_{ab}$ dependent function $|K|dG$ into tensor products of
its vectors. Formally, this splitting parallels the known generalization of the
integrated form of the kinematic fundamental equation of integral geometry
(\ref{split}). But the Minkowski functionals \cite{santalo-book} are now
replaced by the infinite set of 1-point measures 
\begin{equation}\label{measure-1}
\mu^A_k(D)= p_\sigma(\{\kappa_\alpha\})\prod_{i=0}^\infty (e_i)^{\otimes
M_i}\otimes\pi
\end{equation}
for the f-function $A$ at a fixed root point (see FIG.~\ref{fig:vertex}), where
the multi-index $k=(\sigma, \vec{M})$ carries further information about the
polynomial $p_\sigma$ of principal curvatures $\kappa_\alpha$, classified by the
Young diagram $\sigma$ of the symmetric group and the tensorial exponents
$M_i\in\mathbb{N}^*$ of the vectors. The additional variable $\pi$ indicates the
parity of the tensor valued function under axial rotations, defined later. 

This infinite tensor space is the result of an explicit manipulation of
(\ref{vertex-reduced}). But there is one alternative approach that might hint
at connections to index theory \cite{bott-tu}. From the Gauss-Codazzi equation
(\ref{gauss-codazzi}) we see, that $\kappa_\alpha, e_\alpha$ and $e_n$ are
covariantly constant and lie in the kernel of the eigenvalue equation
$D^{(n-1)}\Phi= \lambda\Phi$. Therefore, we can characterize the set of 1-point
measures as the tensor valued space 
\begin{equation}\label{space}
\mathcal{M}(\Sigma, \mathbb{R}^n)=\{\Phi(\kappa_\alpha, e_i)\, |\,
D^{(n-1)}\Phi=0\}
\end{equation}
leading to the second result that {\em the irreducible elements of
$k$-vertices decompose into the infinite dimensional tensor space of 1-point
measures $\mathcal{M}( \Sigma, \mathbb{R}^n)$ with multiplicative structure
defined by the kinematic equation (\ref{vertex-reduced}).} Comparing this to
the Hilbert space of quantum mechanics, one might interpret the 1-point
measures as the wave functions $\phi$ of a massless bosonic particle with
vertex interactions up to order $\phi^n$.

There are many open questions concerning this space, of which some will be
discussed in \cite{korden}. Let us here only mention the case of Riemannian
surfaces $T_g$ of genus $g$ and Euler characteristic $\chi(T_g)=2-2g$. For
$g\geq 1$ the volume dependent part of the second virial $\chi(D_1)
\text{vol}(D_2)$ is either zero or negative, contradicting experience. This
inconsistency follows from the representation in Mayer functions and can be
mended by including further homotopic invariants as the Euler linking number
\cite{bott-tu}. As these forms are again decomposable into 1-point measures,
the above result remains unchanged even for homotopically nontrivial manifolds.

{\em Rebuilding the Cluster ---}
With the decoupling of the cluster integrals into vertices, diagrams as that of
FIG.~\ref{fig:vertex} reduce to expressions of the form
\begin{equation}\label{4-point}
\begin{aligned}
& C^{n\left[AE\right.}_{i_1l_1} C^{\left. ED\right]l_1}_{j_1k_1}
C^{n\,AB}_{j_2i_2}
C^{n \left[BE\right.}_{j_3l_3}C^{\left. EC\right]l_3}_{k_3i_3}
C^{n\,CD}_{j_4i_4}\\
& \times (\mu^A_{i_1}\mu^A_{j_2}) (\mu^B_{i_2}\mu^B_{j_3}) 
(\mu^C_{i_3}\mu^C_{j_4}) (\mu^D_{i_4}\mu^D_{j_1}) (\mu^E_{k_1}\mu^E_{k_3})
\end{aligned}
\end{equation}
where the angular brackets between the coefficients $C^k_{ij}$ indicate the
symmetric permutation of f-labels. Calculating the cluster integral is now
reduced to integrating pair products of 1-point measures $(\mu^A_{i_1}
\mu^A_{j_2})$ for the rigid particle $\Sigma_A$. But both measures have been
derived at independent points in the embedding space, here denoted by
$\vec{r}_1$ and $\vec{r}_2$. It is therefore necessary to merge these two
coordinate systems into the one of the rigid body, denoted by $(\fe_1, \ldots,
\fe_{n})$, with the axial direction $\fe_n r = \vec{r}_1 -\vec{r}_2$.
Generalizing Wertheim's discussion to $n$ dimensions, the tensorial
part of (\ref{measure-1}) can be reduced by integral averaging over axial
rotations, resulting in breaking up the symmetry group $\SO{n}$ into
$\SO{n}/\SO{n-1} \times \SO{n-1} = S^{n-1}\times \SO{n-1}$. But the two Lie
groups $\SO{2l+1} = B_l$ and $\SO{2l} = D_l$ belong to different Dynkin
diagrams, where $D_l$ carries an additional $\mathbb{Z}_2$ automorphism
\cite{helgason} that induces possible sign changes under axial rotations. This
is an additional degree of freedom, we have taken care of by including the
parity symbol $\pi$ in (\ref{measure-1}). The existence of such odd and even
parity tensors might have a significant influence on the solid phase structure,
as the free-energy functional could develop further minima in addition to the
pointwise vanishing of the group measures in (\ref{vertex-reduced}) by
$\sin{(\phi_{ab})}=0$. It would therefore be interesting to compare the odd
to the even dimensional phase structure, with a special focus on $n=8$ and its
$\mathbb{Z}_3$ automorphism group of $D_4$.

Calculating the axial average of products of $\SO{n}$ invariant tensors
$|\lambda, m_1, \ldots, m_l>$, reduces now to determining their $\SO{n-1}$
invariant subspaces
\begin{equation}\label{reduce}
\begin{aligned}
<\lambda, \vec{m}'|B|\lambda, \vec{m}> & = \text{const.} \quad \text{for} \quad
B\in\SO{n-1}\\
\Rightarrow & \quad \Lambda(\vec{m}',\vec{m})=0
\end{aligned}
\end{equation}
represented by linear constrains between their weight vectors $\Lambda(\vec{m}',
\vec{m}) = 0$. A general solution of this problem has been given by Cartan
\cite{helgason} and extended to a complete classification of Riemannian
symmetric spaces. We can therefore assume the constrains $\Lambda=0$ to be 
known and indicate axially symmetric tensors of rank
$\lambda$ and weight vector $\vec{m}$ by $\left|\lambda, \vec{m},
e>\right|_{\Lambda=0}$, where the third component indicates either the
coordinate system of the vertex $e$, the rigid body $\fe$ or that of the
embedding space $E$. Following the notation of \cite{wertheim-1, wertheim-2, 
wertheim-3, wertheim-4}, the 2-point measures are tensor products of the form 
\begin{equation}\label{2-point}
\begin{aligned}
\mu(\vec{r}_i)\otimes \mu(\vec{r}_j)  = & \\
f(r_{ij}, \lambda_i, \lambda_j & ,\vec{M}_i,\vec{M}_j)
|\lambda_i,\vec{M}_i,e_i>\otimes |\lambda_j, \vec{M}_j,e_j>\;,
\end{aligned}
\end{equation}
whose tensorial part will also be written as $|e_i>\otimes |e_j>$ if no further
reference to the representation space is required. Joining the two coordinate
systems $e_i, e_j$ into that of the rigid body $\fe$, corresponds to an
integration
\begin{equation}
\begin{aligned}
\int <\fe| e_i><\fe| e_j> d\omega_i d\omega_j\; &
|\fe> \otimes |\fe> \\
= |\fe> \otimes |\fe>|_{\Lambda=0}\;, &
\end{aligned}
\end{equation}
over the Haar measure $d\omega$ of $\SO{n-1}$, reducing the space by
$(n-1)(n-2)/2$ dimensions. The resulting 2-point measures 
\begin{equation}\label{2-p}
f(r)|\fe> \otimes |\fe>|_{\Lambda=0}
\end{equation}
are now irreducible tensors in the coordinates $(\fe,r) \in
S^{n-1}\times\mathbb{R}$ of the particle $\Sigma$.

The 2-point measures and the expansion coefficient of the vertex (\ref{split})
carry the local properties of the particle $\Sigma$. But the cluster diagrams
also carry global informations, as can be seen from FIG.~\ref{fig:vertex}.
Consider the star graphs of the free-energy. With the exception of the second
virial, all such diagrams are build from closed subdiagrams or loops,
introducing a set of constrains $\vec{r}_{12}+ \ldots+ \vec{r}_{m1} = 0$ between
the coordinates of the $m$ root points of one loop. These constrains also couple
particles that otherwise  do not intersect. This problem occurs first for ring
graphs and has been solved by Wertheim in  \cite{wertheim-1, wertheim-2,
wertheim-3, wertheim-4} by introducing the Radon transformation \cite{gelfand}
to replace the particle fixed coordinate system $(\fe, r)$ by that of the
embedding space $(E,R)$. This approach can be extended in a straightforward
way to arbitrary diagrams in $n$ dimensions. The Radon transformed 2-point
function
\begin{equation}\label{F-function}
\begin{aligned}
F(R,E)  & \\
= \int & <E|\fe> <E|\fe>|_{\Lambda=0} f(r) \delta(r\fe E-R) d^n \mathfrak{r} 
\end{aligned}
\end{equation}
can then be seen as the propagator of a state vector $|E>$ along a line of
distance $R$. With $\left|E>F(R,E)<E\right|$ interpreted as the Green's
function, the Mayer cluster can be seen as the analog of Feynman diagrams,
with the constrains 
\begin{equation}\label{charges}
(r_{12}\fe_{12} + \ldots + r_{m,1} \fe_{m,1})E = 0\;, \quad  
\Lambda = 0
\end{equation}
as the conservation laws for each loop. Calculating the virial coefficient
reduces now to an integration over the embedding space coordinates $(R,E)$,
completing the argument that arbitrary hard-particle clusters in $n>1$
dimensions decouple into 2-point measures. The advantage of this approach
is not an efficient calculational tool but the detailed connection between the
1-point measures and their occurrence in the free-energy, opening up a path to
better understanding the phase structure of hard-particle fluids by the
particle geometry and relative orientations.

{\em Acknowledgment ---}
This work was performed as part of the Cluster of Excellence "Tailor-Made Fuels
from Biomass", funded by the Excellence Initiative of the German federal and
state governments.
%
\bibliography{Short_Proof_for_Reducibility}
\end{document}